\documentclass[aps,showpacs,nofootinbib,superscriptaddress]{revtex4}
\usepackage{units}
\usepackage{graphicx,subfigure}  
\usepackage{amsmath} %besserer Mathesatz
\usepackage{amsfonts} %bestimmte Symbole
\usepackage{amssymb} %noch mehr bestimmte Symbole
\usepackage{slashed} %Feynman slash notation

\begin{document}
\title{In-medium properties of the $\phi$ meson with $\phi~N$ resonant contributions}

\author{D. \surname{Cabrera}}
\affiliation{Departamento de F\'\i sica Te\'orica and IFIC, Centro Mixto
Universidad de Valencia-CSIC, Institutos de Investigaci\'on de
Paterna, E-46071 Valencia, Spain}
\author{A. N. \surname{Hiller Blin}}
\affiliation{Departamento de F\'\i sica Te\'orica and IFIC, Centro Mixto
Universidad de Valencia-CSIC, Institutos de Investigaci\'on de
Paterna, E-46071 Valencia, Spain}
\author{M. J.  \surname{Vicente Vacas}}
\affiliation{Departamento de F\'\i sica Te\'orica and IFIC, Centro Mixto
Universidad de Valencia-CSIC, Institutos de Investigaci\'on de
Paterna, E-46071 Valencia, Spain}

\today

\begin{abstract}
Nuclear production experiments report missing absorption processes of the in-medium $\phi$ meson. Contributions arising from the $\bar KK$ cloud have already been widely studied, and therefore we investigate the $\phi$-meson's properties in cold nuclear matter with the additional inclusion of resonant $\phi~N$ interactions. Two models are considered which dynamically generate $N^*$-like states close to the $\phi~N$ threshold. We find that these states, together with the non-resonant part of the amplitude, contribute to the $\phi$ self-energy with the same order of magnitude as the $\bar KK$ effects. At non-vanishing nuclear density, both models lead to an additional in-medium broadening of the $\phi$, up to around $50~\unit{MeV}$. Furthermore, at least one of the models is compatible with a mass shift to lower energies of up to $35$~MeV at threshold and normal matter density. Finally, a double-peak structure appears in the spectral function due to the mixing of resonance-hole modes with the $\phi$ quasi-particle peak. These results converge into the direction of the experimental findings.
\end{abstract}


\maketitle
\section{Introduction}

The $\phi$ production in nuclei has been intensively studied both experimentally~\cite{Muto:2005za,Ishikawa:2004id,Wood:2010ei,Hartmann:2012ia}, as well as in theoretical analyses~\cite{Cabrera:2003wb,Magas:2004eb,Hartmann:2012ia,Paryev:2008ck,Muhlich:2005kf}. The $\phi$ nuclear transparency ratio measured in proton-induced and photoproduction experiments points to a large absorption in nuclei, consistent with widths about one order of magnitude larger than the one from its dominant decay to $\bar K K$ in vacuum. In Ref.~\cite{Wood:2010ei} widths of $23-100$~MeV are reported, even for densities of only half the normal nuclear-matter density. Since the measurement of $\phi$ absorption with momentum binning has been made~\cite{Hartmann:2012ia}, theoretical calculations of the momentum dependence of the $\phi$ self-energy are also of great interest.

Heavy-ion collision (HIC) experiments, too, have been investigating the production of $\phi$ mesons~\cite{Wada:2013mua,Abelev:2014uua}. They observed a deep sub-threshold $\phi$ production~\cite{Lorenz:2014eja}. A possible explanation for this observations might be a change in the $\phi$-meson properties due to, for example, an attractive mass shift and an enhanced decay width arising from $\phi~N$ interactions. This is further supported by the large absorption seen in previous experiments.

Due to its hidden strangeness content, the $\phi$ strongly couples to the $\bar K K$ system. Therefore, also the in-medium dynamics are expected to be mainly governed by its decay into these light pseudoscalars. In fact, the direct coupling of the $\phi$ to the nucleon is OZI-forbidden. Nevertheless, the above mentioned experimental results require medium effects larger than predicted by this kind of hadronic models, and additional interactions with the nuclear medium should be taken into account. Therefore, it is important to study possible interaction and production mechanisms that may have been missed in theoretical analyses so far.

The $\phi$ self-energy originating from medium effects on the $\bar K K$ decay mode has been broadly studied in the past, both at zero and non-vanishing temperatures, using chiral-dynamic constraints on the pseudoscalar-meson interactions with nucleons and light mesons~\cite{Ko:1992tp,Klingl:1997tm,Holt:2004tp,AlvarezRuso:2002ib,Oset:2000eg,Cabrera:2002hc,Faessler:2002qb}. While the modification of the $\phi$ width is large, it does not entirely achieve to explain the experimental observation.

Here, we revisit the $\phi$ self-energy in cold nuclear matter by analyzing the effect of direct $\phi$-nucleon interactions not previously considered. The hidden strangeness in the $\phi$ meson can be indirectly exchanged with the in-medium nucleon by the coupling to $K^*\Lambda$ and $K^* \Sigma$ pairs, without violation of the OZI rule. Elastic $\phi$ scattering off the nucleon is then possible via $K^*Y$ loops, which can be studied in unitary coupled-channel approaches. In these models, several broad $N^*$ states are dynamically generated around energies close to the $\phi~N$ threshold. Furthermore, they have a non-negligible coupling to this channel, therefore being good candidates for the enhancement of the $\phi$ self-energy. We explore whether these resonant $\phi~N$ interactions can fill in for the missing absorption pointed out by nuclear production experiments.

In Sec.~\ref{sec:model}, we briefly show the formalism for the calculation of the $\phi$ self-energy with $\phi~N$-interaction mechanisms. We present and discuss our results for the $\phi$ nuclear optical potential and spectral function in Sec.~\ref{sec:results}.

\section{Formalism}\label{sec:model}

Since the $\bar K K$-cloud contributions to the $\phi$-meson self-energy have already been thoroughly studied in literature, here we want to concentrate on the novel in-medium processes, namely the presence of resonances in the $\phi~N$ scattering amplitude in the vicinity of the $\phi~N$ threshold. To estimate the uncertainties, we compare two approaches to this implementation, since they differ in size and energy dependence: In Refs.~\cite{Oset:2009vf,Oset:2012ap}, vector mesons are introduced through the Hidden Local Symmetry (HLS) approach, where the tree level vector-meson nucleon scattering amplitudes are obtained from a vector-meson
exchange mechanism; In Ref.~\cite{Gamermann:2011mq}, it is obtained within an SU(6) spin-flavor symmetry extension of chiral perturbation theory.

The $\phi$ self-energy is obtained by integrating the scattering amplitude $T$ over the nucleon Fermi distribution,
\begin{equation}
\label{eq:phi-N-self}
\Pi^{\phi~N}_{\phi}(q^0,\vec{q};\rho) =
4 \int \frac{d^3 p}{(2\pi)^3} n(\vec{p}) \frac{1}{6}[2 T^{\frac{1}{2}\frac{1}{2}}_{\phi~N}(P^0,\vec{P})+4 T^{\frac{1}{2}\frac{3}{2}}_{\phi~N}(P^0,\vec{P})] \ ,
\end{equation}
where $P^0=q^0+E_N(\vec{p})$ and $\vec{P}=\vec{q}+\vec{p}$ are the total energy and momentum of the $\phi~N$ pair in the nuclear-matter rest frame. The scattering amplitude in the two models has to be calculated by solving the coupled-channel on-shell Bethe-Salpeter equation, $T = [1-VG]^{-1} V$. For further definitions, refer to our work in Ref.~\cite{Cabrera:2016rnc}. Additionally, we took into account the modifications of $G$ from Pauli blocking. This function then reads
\begin{eqnarray}
\label{eq:Pauli}
G(P;\rho) &=& G(P) + \delta G^{\rm Pauli}(P,\rho) \ , \nonumber \\
\delta G^{\rm Pauli}(P,\rho) &=& \int \frac{d^3 q}{(2\pi)^3} \frac{M_N}{E_N(\vec{p})} \frac{-n(\vec{p})}{[P^0-E_N(\vec{p})]^2-\omega_{\phi}^2(\vec{q})+i\epsilon}\bigg |_{\vec{p}=\vec{P}-\vec{q}} \ ,
\end{eqnarray}
with $\omega^2_{\phi}(\vec{q})=M_{\phi}^2+\vec{q}\,^2$ and $n(\vec{p})$ the Fermi-gas nucleon momentum distribution, $n(\vec{p})=\Theta(p_F-|\vec{p}|)$. The Fermi momentum is given in terms of the nuclear-matter density $\rho$ by $p_F=(3\pi^2\rho/2)^{1/3}$. We leave the consideration of additional sources of medium effects, such as baryon binding potentials and the self-consistent dressing of the $\phi$ propagator within the $\phi~N$ loop function, to future works.

\section{Results and discussion}
\label{sec:results}

\begin{figure}[t]
\centering
\includegraphics[width=0.42\textwidth]{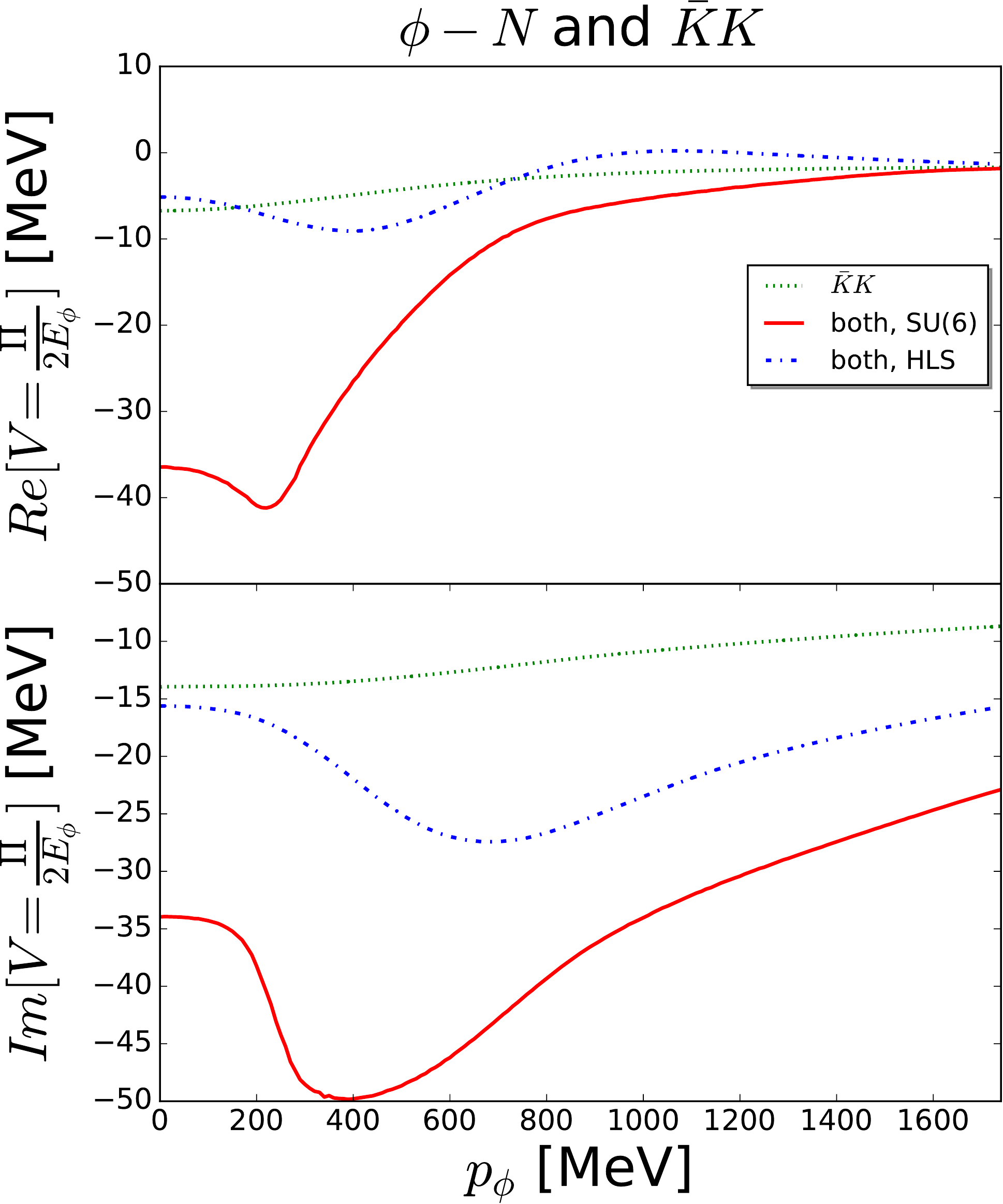} \hspace{10mm}
\includegraphics[width=0.495\textwidth]{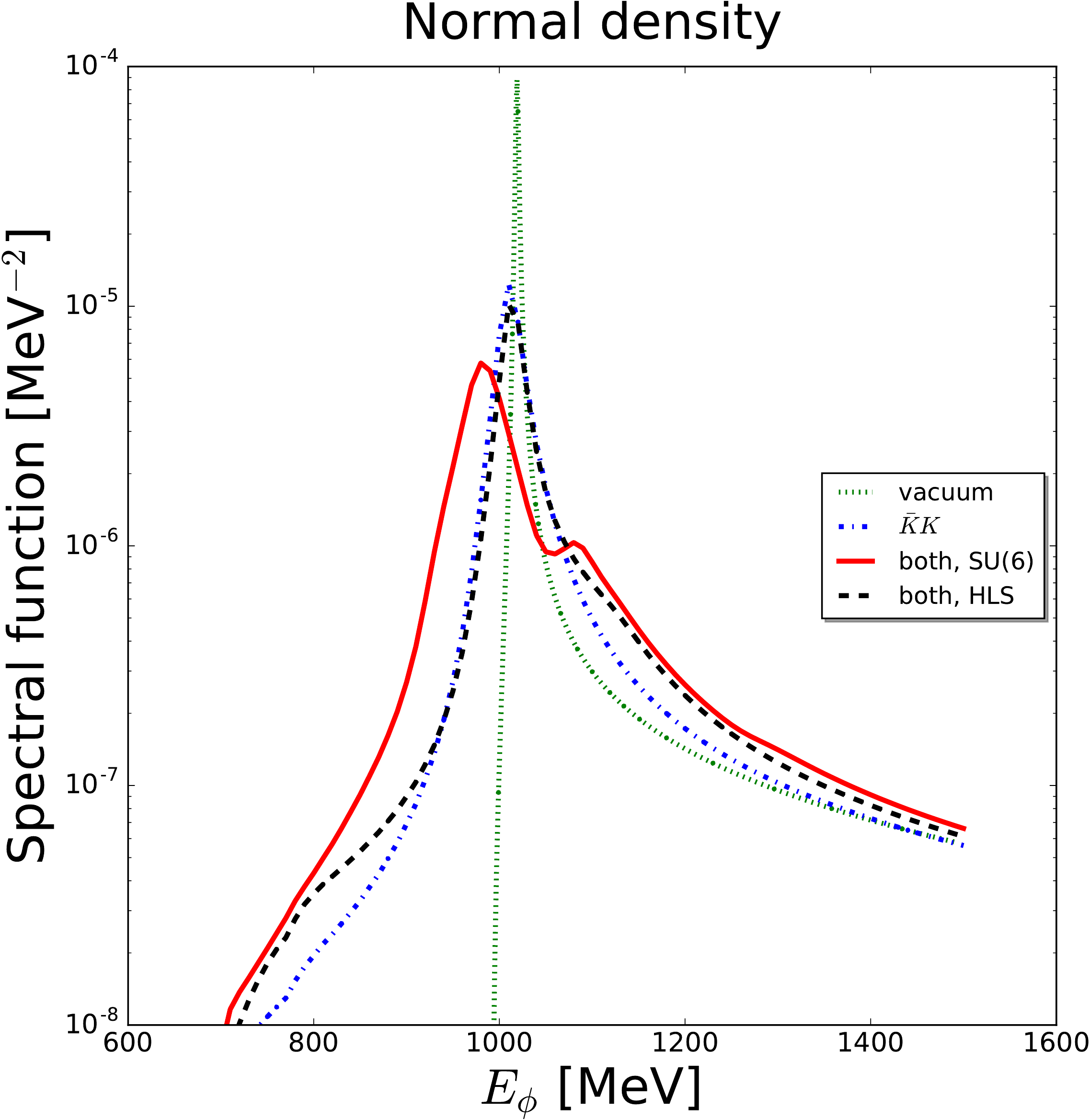}
\caption{Left: Real and imaginary parts of the $\phi$ optical potential as a function of the $\phi$ momentum at normal matter density. Contribution from the $\bar K K$ cloud (dotted) and the coherent sum including $\bar K K$ plus $\phi~N$ effects in the SU(6) (solid) and HLS (dash-dotted) approaches. Right: $\phi$ spectral function as a function of the (off-shell) $\phi$ energy at zero momentum. Calculation accounting for the $\bar K K$ self-energy (dash-dotted) or the $\bar K K$ plus $\phi~N$ self-energies in the HLS model (dashed) and the SU(6) model (solid), all three cases considered at $\rho=\rho_0$. The $\phi$ spectral function in vacuum is also shown (dotted).}
\label{fig:Vopt}
\end{figure}

On the left panel of Fig.~\ref{fig:Vopt}, we show the $\phi$ nuclear optical potential at normal nuclear density, as a function of the $\phi$ momentum. We take into account the contributions from the $\bar K K$-cloud and the $\phi~N$ interactions. These two mechanisms give rise to imaginary parts --- related to the width enhancement --- of the same order of magnitude. Concerning the $\phi~N$ interactions, the magnitude at zero momentum is larger in the SU(6) than in the HLS model. Nevertheless, even the latter is substantial for typical production-experiment momenta around $500~\unit{MeV}$. The real part, which can be connected to a mass shift, is of importance only when calculated in the SU(6) approach.

On the right panel of Fig.~\ref{fig:Vopt}, we depict the spectral function of an off-shell $\phi$ meson at vanishing momentum and normal density, as a function of its energy. Again, the case where the $\bar K K$ self-energy alone is considered is compared with the cumulative contribution of both the $\bar K K$ cloud and the $\phi~N$ interactions. The latter are shown in the HLS approach, as well as in the SU(6) model. For reference, the $\phi$ spectral function in vacuum is shown, too. As has been shown in other works, the effects of the $\bar K K$ cloud lead to a broadening of the $\phi$ spectral function, and to an enhancement of the low-energy tail in nuclear matter~\cite{Klingl:1997tm,Oset:2000eg,Cabrera:2002hc}. The mass shift is negligible. When adding the $\phi~N$-interaction contributions, there is an increase of the broadening, and an enhancement of the low-energy tail. Furthermore, a second shoulder above the $\phi$ mass appears due to the excitation of resonance-hole modes. In the SU(6) model, the $\phi$ quasi-particle peak is shifted to lower energies in agreement with the observed attractive optical potential.

In summary, the absorption of the $\phi$ meson in matter suffers a strong and energy-dependent enhancement from resonant $\phi~N$ interactions. This is in agreement with the results obtained in nuclear production experiments. As a next step, it is important to implement this updated self-energy in order to analyse the nuclear transparency ratio for different nuclei, and as function of $\phi$ momenta. Furthermore, the $\phi$ production reaction in nuclei could be used to check the validity of the HLS and SU(6) models, due to their differing results.

\section*{Acknowledgements}
We acknowledge fruitful discussions with Elena Bratkovskaya, Wolfgang Cassing, Juan Nieves, Eulogio Oset, \`Angels Ramos and Laura Tol\'os. This research has been partially supported by the Spanish Ministerio de Econom\'{\i}a y Competitividad (MINECO) and the European fund for regional development (FEDER) under Contracts FIS2011-28853-C02-01, FIS2014-51948-C2-1-P, FIS2014-51948-C2-2-P and SEV-2014-0398, by the Helmholtz International Center for FAIR within the framework of the LOEWE program, and by Generalitat Valenciana under Contract PROMETEOII/2014/0068. A.N.~Hiller Blin acknowledges support from the Santiago Grisol\'{\i}a program of the Generalitat Valenciana. D.~Ca\-bre\-ra acknowledges support from the BMBF (Germany) under project No.~05P12RFFCQ.

\end{document}